\documentstyle[preprint,aps,floats,cite,psfig,subeqn]{revtex}
\begin{document}


\draft

\preprint{\vbox{\baselineskip14pt\hbox{\bf KEK-TH-440}
\hbox{\bf KEK Preprint 95-42}\hbox{\bf KANAZAWA-95-06}\hbox{\today}}}

\title{Constraints on New Physics in the Electroweak Bosonic \\
             Sector from Current Data and Future Experiments}

\author{K.~Hagiwara$^1$, S.~Matsumoto$^{1,2}$ and R.~Szalapski$^1$}
\address{$^1$Theory Group, KEK, Tsukuba, Ibaraki 305, Japan \\
	 $^2$Department of Physics, Kanazawa University, Kanazawa 920-11, Japan}

\maketitle


\begin{abstract}

Extensions of the Standard Model which involve a new scale, $\Lambda$, may,
for energies sufficiently small compared to this new scale, be expressed in
terms of operators with energy dimension greater than four.  The coefficients
of just four SU(2)$\times$U(1)-gauge-invariant energy-dimension-six
operators are sufficient to parameterize the contributions of
new physics in the electroweak bosonic sector to electroweak precision
measurements. In this letter we update constraints on the coefficients of
these four operators due to recent precision measurements of electroweak
observables.  We further demonstrate how such constraints may be improved  by
experiments at TRISTAN, LEP2 and at a future linear $e^+e^-$ collider.
The relationship of these operators to the oblique
parameters  $S$, $T$ and $U$ is examined.  Two of the operators
contribute to a non-standard running of the electroweak charge form-factors
$\overline{\alpha}(q^2)$, $\overline{s}^2(q^2)$,
$\overline{g}_Z^2(q^2)$ and $\overline{g}_W^2(q^2)$;  in the special case
where the coefficients of these two vanish the operator analysis
reduces to an analysis in terms of  $S$, $T$ and $U$ with $U = 0$.

\end{abstract}
\newpage


If one assumes that the Standard Model (SM) is the low-energy approximation of
some more general theory for which we lack a complete description, then it is
possible to describe the full theory with an effective Lagrangian.  The leading
terms of the effective Lagrangian will be the energy-dimension-four
operators of
the SM, and correction terms will be operators of greater energy dimension
suppressed by inverse powers of the scale of the additional physics, $\Lambda$.
Under the assumption $\Lambda \gg v$, where $v = 246.22$GeV
is the vacuum expectation value (vev) of the SM Higgs field, one may write
an effective Lagrangian of the full theory in the form
\begin{equation}\label{fulllagrangian}
{\cal L}_{eff} = {\cal L}_{\rm SM}
+  \sum_{n \geq 5} \sum_i \frac{f_i^{(n)} \,
{\cal O}_i^{(n)}}{\Lambda^{n-4}} \;.
\end{equation}
The energy dimension of each operator is denoted by $n$, and the index $i$ sums
over all operators of the given energy dimension.
Assuming that the correct low-energy theory is the SM and that the full
theory is SU(2)$\times$U(1) gauge invariant, these
operators are constructed from SM fields and derivatives.  An exhaustive list
of energy-dimension-five and -six operators has been compiled in
Refs.~\cite{BW,bsllr}.

By studying the consequences of these operators one may hope to acquire
model-independent insight into the new physics.  In general,
however, even if the analysis is restricted to operators not exceeding
energy-dimension-six, so many operators contribute that the effective
Lagrangian
lacks predictive power.  Nevertheless there is a special class of new physics
which may be effectively studied through the present precision electroweak
measurements.  All physics which contributes to precision measurements only
via contributions to the two-point functions of SM gauge bosons belongs to this
class \cite{LPS,STU,eps123,others}.
It was first observed by Grinstein and Wise \cite{GW} that only four operators
through energy-dimension-six
contribute at the tree level.
In the notation of Ref.~\cite{HISZnew} they are
\begin{subequations}
\label{four_operators}
\begin{eqnarray}
\label{odw}
{\cal O}_{DW} & = & - g^2 {\rm Tr} \bigg[ \Big(\partial_\mu W_{\nu\rho} \Big)
\Big( \partial^\mu
W^{\nu\rho}\Big)\bigg]\; ,  \\
\label{odb}
{\cal O}_{DB} & = & -\frac{g'^2}{2}\Big(\partial_\mu
B_{\nu\rho}\Big)\Big(\partial^\mu B^{\nu\rho}\Big)\; , \\
\label{obw}
{\cal O}_{BW} & = & - \frac{g g^\prime}{2}
   \Phi^\dagger B_{\mu\nu}W^{\mu\nu} \Phi\; , \\
\label{ophi1}
{\cal O}_{\Phi,1} & = & \bigg[ \Big(D_\mu\Phi\Big)^\dagger \Phi\bigg] \;
\bigg[ \Phi^\dagger \Big(D^\mu \Phi\Big)\bigg]\;.
\end{eqnarray}
\end{subequations}
Here $W_{\mu\nu} = W_{\mu\nu}^a T^a$, $g$ is the SU(2)
coupling with ${\rm Tr}(T^aT^b)=\frac{1}{2}\delta^{ab}$,
and $g^\prime$ is the U(1) coupling.
Their coefficients, the $f_i^{(n)}$ of Eq.~(\ref{fulllagrangian}),
are denoted by
\begin{equation}\label{fbasis}
\{f_{DW}, f_{DB}, f_{BW}, f_{\Phi,1}\}
\end{equation}
respectively.


{\bf Oblique corrections:}
To date precision electroweak measurements are restricted to those processes
which involve four external light fermions and conserve both chirality and
flavor.  Those effects of the new physics which may be described by the
operators~(\ref{four_operators}) contribute to electroweak
precision measurements
via their contributions to the transverse components of the gauge-boson
propagators.  If the one-particle-irreducible two-point-function
is separated into SM and new physics contributions according to
$ \Pi = \Pi_{\rm SM} + \Delta\Pi $, then, in the notation
of Ref.~\cite{HISZnew},
we find
\begin{subequations}
\label{two_point_functions}
\begin{eqnarray} \label{piqq}
\Delta\Pi^{QQ}_T(q^2) &=&  2 \frac{q^2}{\Lambda^2} \biggl[ (f_{DW}+ f_{DB}) q^2
 - f_{BW} \frac{v^2}{4} \biggr]\; ,
\\ \label{pi3q}
\Delta\Pi^{3Q}_T(q^2) &=& 2 \frac{q^2}{\Lambda^2} \biggl[ f_{DW} q^2
 - f_{BW} \frac{v^2}{8} \biggr]  \; ,
\\ \label{pi33}
\Delta\Pi^{33}_T(q^2) &=& 2 \frac{q^2}{\Lambda^2} f_{DW} q^2
 -  \frac{v^2}{\Lambda^2} f_{\Phi,1} \frac{v^2}{8} \; ,
\\ \label{pi11}
\Delta\Pi^{11}_T(q^2) &=& 2 \frac{q^2}{\Lambda^2} f_{DW} q^2 \; .
\end{eqnarray}
\end{subequations}
One immediately notices that these expressions contain terms which
are constant and linear in $q^2$, but terms which are quadratic in $q^2$
appear as well \cite{GW,HISZnew}.  For precisely this reason the
dimension-six operator analysis does not reduce to the standard analysis in
terms of a triplet of parameters such as $S$, $T$ and $U$ of Ref.~\cite{STU},
$\epsilon_1$, $\epsilon_2$ and $\epsilon_3$ of Ref.~\cite{eps123} or others of
Ref.~\cite{others}.  By contrast the standard three-parameter description is
sufficient for an analysis based upon the effective Lagrangian with a
non-linear realization of the gauge symmetry.
(See Ref.~\cite{AW} and references
therein.)

It is convenient to introduce an additional parameter for which we choose
the anomalous contribution to the running of $\alpha_{QED}$ evaluated at
$m_Z^2$.  We define
$\Delta\frac{1}{\alpha} \equiv \Delta\overline{\alpha}^{-1}(m_Z^2)$.
Then we may form a quartet of parameters given by
\begin{equation}\label{STUbasis}
\{\Delta S,\Delta T,\Delta U, \Delta\frac{1}{\alpha}\},
\end{equation}
where $S=S_{\rm SM} + \Delta S$, $T=T_{\rm SM} + \Delta T$ and
$U=U_{\rm SM} + \Delta U$.
The quartet (\ref{STUbasis}) is related to (\ref{fbasis}) according to
\begin{subequations}
\label{relationship}
\begin{eqnarray}\label{deltas}
\Delta S & \equiv & 16\pi\,{\cal R}\!e
\bigg[\Delta\Pi^{3Q}_{T,\gamma}(m^2_Z) - \Delta\Pi^{33}_{T,Z}(0)\bigg]
= - 4\pi \frac{v^2}{\Lambda^2}f_{BW}
\;,
\\ \label{deltat}
\Delta T & \equiv & \frac{4\sqrt{2} G_F}{\alpha} \,{\cal R}\!e
\bigg[\Delta\Pi^{33}_{T}(0) - \Delta\Pi^{11}_{T}(0)\bigg]
= - \frac{1}{2 \alpha} \frac{v^2}{\Lambda^2}f_{\Phi,1}
\;,
\\ \label{deltau}
\Delta U & \equiv & \makebox[0.12cm]{}
 16\pi\,{\cal R}\!e
\bigg[\Delta\Pi^{33}_{T,Z}(0) - \Delta\Pi^{11}_{T,W}(0)\bigg]
\makebox[0.13cm]{}
= 32\pi \frac{m_Z^2-m_W^2}{\Lambda^2}f_{DW}
\;,
\\ \label{deltaalpha}
\Delta\frac{1}{\alpha} & \equiv & \makebox[0.12cm]{}
 4\pi\,{\cal R}\!e
\bigg[\Delta\Pi^{QQ}_{T,\gamma}(m^2_Z) - \Delta\Pi^{QQ}_{T,\gamma}(0)\bigg]
\makebox[0.13cm]{} =
8\pi \frac{m_Z^2}{\Lambda^2} \Big(f_{DW} + f_{DB}\Big)
\;.
\end{eqnarray}
\end{subequations}
where the following short-hand notation\cite{HHKM} has been employed:
\begin{eqnarray}\label{defn-pitv}
\Pi^{AB}_{T,V}(q^2) = {\Pi^{AB}_T(q^2) - \Pi^{AB}_T(m^2_V)\over q^2 -
m^2_V}\;.
\end{eqnarray}

Employing (\ref{relationship}) one may calculate the charge form-factors of
Ref.~\cite{HHKM},
\begin{subequations}
\label{gzb20_sb2mz2_gwb20}
\begin{eqnarray}
\label{gzb20}
\frac{1}{ \overline{g}_Z^2(0) }& = & \frac{ 1 + \overline{\delta_G}
- \alpha T}{ 4 \sqrt{2}
G_F m_Z^2}\;, \\
\label{sb2mz2}
\overline{s}^2(m_Z^2) & = & \frac{1}{2} - \sqrt{ \frac{1}{4} -
\overline{e}^2(m_Z^2)\left( \frac{1}{ \overline{g}_Z^2(0) }
+ \frac{S}{16\pi}\right)}\;, \\
\label{gwb20}
\frac{1}{ \overline{g}_W^2(0) }& = & \frac{\overline{s}^2(m_Z^2)}{
\overline{e}^2(m_Z^2)} - \frac{1}{16\pi} \left(S+U\right)\;,
\end{eqnarray}
\end{subequations}
which are intimately related to observable quantities.
The SM vertex and box corrections to the muon lifetime
are incorporated in $\overline{\delta_G} \approx 0.0055$.
The remaining effects of the operators (\ref{four_operators})
are expressed as a non-standard running of the charge form-factors:
\begin{subequations}
\label{running}
\begin{eqnarray}
\label{run_alpha}\Delta\Bigg[
\frac{1}{\overline{e}^2(q^2)} - \frac{1}{4\pi\alpha} \Bigg]&=&
2 \frac{q^2}{\Lambda^2} \Big(f_{DW} + f_{DB}\Big)\;,\\
\label{run_sb2}\Delta\Bigg[
\frac{\overline{s}^2(q^2)}{\overline{e}^2(q^2)}
- \frac{\overline{s}^2(m_Z^2)}{\overline{e}^2(m_Z^2)}\Bigg] & = &
2 \frac{q^2 - m_Z^2}{\Lambda^2} f_{DW}\;,\\
\label{run_gzb2}\Delta\Bigg[
\frac{1}{ \overline{g}_Z^2(q^2) } - \frac{1}{ \overline{g}_Z^2(0) }\Bigg] & = &
2 \frac{q^2}{\Lambda^2}\Big( \hat{c}^4 f_{DW} + \hat{s}^4 f_{DB} \Big)\;,\\
\label{run_gwb2}\Delta\Bigg[
\frac{1}{ \overline{g}_W^2(q^2) } - \frac{1}{ \overline{g}_W^2(0) }\Bigg] & = &
2 \frac{q^2}{\Lambda^2}f_{DW}\;.
\end{eqnarray}
\end{subequations}
In the limit where $f_{DB} = f_{DW} = 0$ all non-standard contributions to the
running of the charge form-factors vanish, and the operator analysis is
equivalent to
an analysis in terms of $S$, $T$ and $U$ with $U = 0$.

Combining Eq.~(\ref{gzb20_sb2mz2_gwb20})
with Eq.~(\ref{running})
the results may be stated compactly as
\begin{subequations}
\label{corrections}
\begin{eqnarray}\label{crctn_alpha}
\Delta \overline{\alpha}(q^2) & = &
      - 8 \pi \hat{\alpha}^2
	\frac{q^2}{\Lambda^2}\Big( f_{DW} + f_{DB} \Big)\;,
\\ \label{crctn_gzbar2}
\Delta \overline{g}_Z^2(q^2) & = &
  	- 2 \hat{g}_Z^4 \frac{q^2}{\Lambda^2} \Big(\hat{c}^4 f_{DW}
	+ \hat{s}^4 f_{DB} \Big)
 	- \frac{1}{2} \hat{g}_Z^2 \frac{v^2}{\Lambda^2} f_{\Phi,1} \;,
\\ \nonumber
\Delta \overline{s}^2(q^2) & = &
	 \frac{-\hat{s}^2\hat{c}^2}{\hat{c}^2-\hat{s}^2}\Bigg[
	8\pi\hat{\alpha}\frac{m_Z^2}{\Lambda^2}\Big( f_{DW} + f_{DB} \Big)
	+ \frac{m_Z^2}{\Lambda^2} f_{BW}
	- \frac{1}{2}\frac{v^2}{\Lambda^2}f_{\Phi,1}
	\Bigg]
\\
\label{crctn_sbar2}
   	&& \makebox[5.5cm]{} + 8\pi\hat{\alpha}\frac{q^2-m_Z^2}{\Lambda^2}
   	  \Big( \hat{c}^2 f_{DW} - \hat{s}^2 f_{DB}\Big)\;,
\\ \label{crctn_gwbar2}
\Delta \overline{g}_W^2(q^2) & = &
	-8\pi\hat{\alpha}\hat{g}^2 \frac{m_Z^2}{\Lambda^2}f_{DB}
	- \hat{g}^2 \frac{ \Delta \overline {s}^2(m_Z^2)}{\hat{s}^2}
	- \frac{1}{4} \hat{g}^4 \frac{v^2}{\Lambda^2}f_{BW}
	- 2\hat{g}^4 \frac{q^2}{\Lambda^2}f_{DW} \;.
\end{eqnarray}
\end{subequations}
The `hatted' couplings satisfy the tree-level relationships
$\hat{e} \equiv \hat{g}\hat{s} \equiv \hat{g}_Z\hat{s}\hat{c}$ and
$\hat{e}^2 \equiv 4\pi\hat{\alpha} $.
For numerical results we use
$\hat{\alpha}=128.72$ and $\hat{s}^2 = 0.2312$.


{\bf Low-energy constraints:}
In the calculation of electroweak observables
we use the vertex and box corrections of the SM, but the charge form-factors of
the SM are modified according to (\ref{corrections}).
We then perform a $\chi^2$ analysis of all available electroweak data
to place constraints upon the four coefficients of
(\ref{fbasis}) for $\Lambda = 1$~TeV.
Included in the analysis are the recent LEP data
\cite{lep_data} and the SLC measurement of the left-right asymmetry
\cite{slc_data}.
We present the central values and one-sigma uncertainties of the four
coefficients along with the correlation matrix:
\begin{eqnarray}
\label{fit_now}
\left.
\begin{array}{lll}
f_{DW} & = & -0.35 + 0.012 \ln x_H - 0.14 x_t \pm 0.62 \\[1mm]
f_{DB} & = & -11 \pm 11 \\[1mm]
f_{BW} & = & 3.1  + 0.072 \ln x_H \pm 2.6 \\[1mm]
f_{\Phi,1} & = & 0.23 - 0.031 \ln x_H + 0.36 x_t  \pm 0.17
\end{array}
\right.
\makebox[2mm]{}
\left(
\begin{array}{lddd}
\dec 1.   & \dec $-$0.323 & \dec    0.151 & \dec $-$0.228  \\[1mm]
          & \dec    1.    & \dec $-$0.979 & \dec $-$0.806  \\[1mm]
          &               & \dec    1.    & \dec    0.905  \\[1mm]
          &               &               & \dec    1.
\end{array} \right)
\end{eqnarray}
where
\begin{equation}
x_t = \frac{m_t-175{\rm GeV}}{100{\rm GeV}}\;,  \makebox[1cm]{}
x_H = \frac{m_H}{100{\rm GeV}} \;.
\end{equation}
The parameterization is good to better than 2\% of the one-sigma errors
in the range $140{\rm GeV}<m_t<220{\rm GeV}$ and
$60{\rm GeV}<m_H<800{\rm GeV}$.

For all four parameters the dependencies upon $m_H$ and $m_t$ arise
from SM contributions only.  For $m_t=175$GeV and $m_H=100$GeV we find
$f_{BW}=3.1\pm 2.6$ and $f_{\Phi,1}=0.23\pm0.17$.
A value of $f_{\Phi,1}$ which is more
consistent with zero indicates a lower value for $m_t$ and/or a larger value
for
$m_H$.  In particular $f_{\Phi,1}$ and $m_t$ are correlated via the linear
contribution of the former and the quadratic contribution of the latter to the
$T$ parameter.  The dependence of $f_{BW}$ upon $m_H$ is weak, and the
dependence on $m_t$ is negligible,
hence it is difficult to make $f_{BW}$ more consistent with zero.

We note, however, that $f_{DB}$ is not well constrained by the present data,
and
there are strong correlations among the coefficients $f_{DB}$, $f_{BW}$ and
$f_{\Phi,1}$. The accurate measurement of $\overline{s}^2(m_Z^2)$
from asymmetries leads to a strong
$f_{DB}$-$f_{BW}$ anti-correlation as well as an $f_{BW}$-$f_{\Phi,1}$
correlation.  The precise measurement of the Z-boson width tightly constrains
$\overline{g}_Z^2(m_Z^2)$, which in turn favors a strong
$f_{DB}$-$f_{\Phi,1}$ anti-correlation.  The measurement of
$\overline{g}_W^2(0)$, via the measurement of $m_W$, strengthens the
$f_{BW}$-$f_{\Phi,1}$ correlation.  See Eqn.~(\ref{corrections}).
The individual measurements produce
correlations between $f_{DW}$ and the other parameters, but these effects tend
to cancel in the overall analysis.

We next add the constraint $f_{DW} = f_{DB} = 0$ and perform a two-parameter
fit
where $f_{BW}$ and $f_{\Phi,1}$
are free parameters.  The results are given
by
\begin{eqnarray}
\label{fit_now_2}
\left.
\begin{array}{lll}
f_{BW} & = & -0.04  + 0.11 \ln x_H \pm 0.28 \\[1mm]
f_{\Phi,1} & = & 0.00 - 0.027 \ln x_H + 0.33 x_t  \pm 0.05
\end{array}
\right.
\makebox[6mm]{}
\left(
\begin{array}{ld}
\dec 1.   & \dec 0.845 \\[1mm]
          & \dec 1.    \\
\end{array} \right)
\end{eqnarray}
Employing (\ref{relationship}) the fit (\ref{fit_now_2}) may be rewritten in
terms of a fit of $S$ and $T$ with $U$ constrained to zero, and we find good
agreement with the results of Ref.~\cite{seiji}.  Comparing
(\ref{fit_now_2}) with (\ref{fit_now}) we notice several common features with
regard to the parameters $f_{BW}$ and $f_{\Phi,1}$;
both fits show a strong $f_{BW}$-$f_{\Phi,1}$ correlation, and both fits
possess similar dependence upon $m_t$ and $m_H$.  However,
in the two-parameter fit the errors are significantly reduced, and both
parameters are completely consistent with zero.  The precise measurement of
$f_{BW}$ and $f_{\Phi,1}$, or, equivalently, the precise measurement of $S$
and $T$, is subject to the assumption that $f_{DW}$ and $f_{DB}$, which are
poorly constrained by the data, are small.

Fig.~1(a)
\begin{figure}[htbp]
\psfig{file=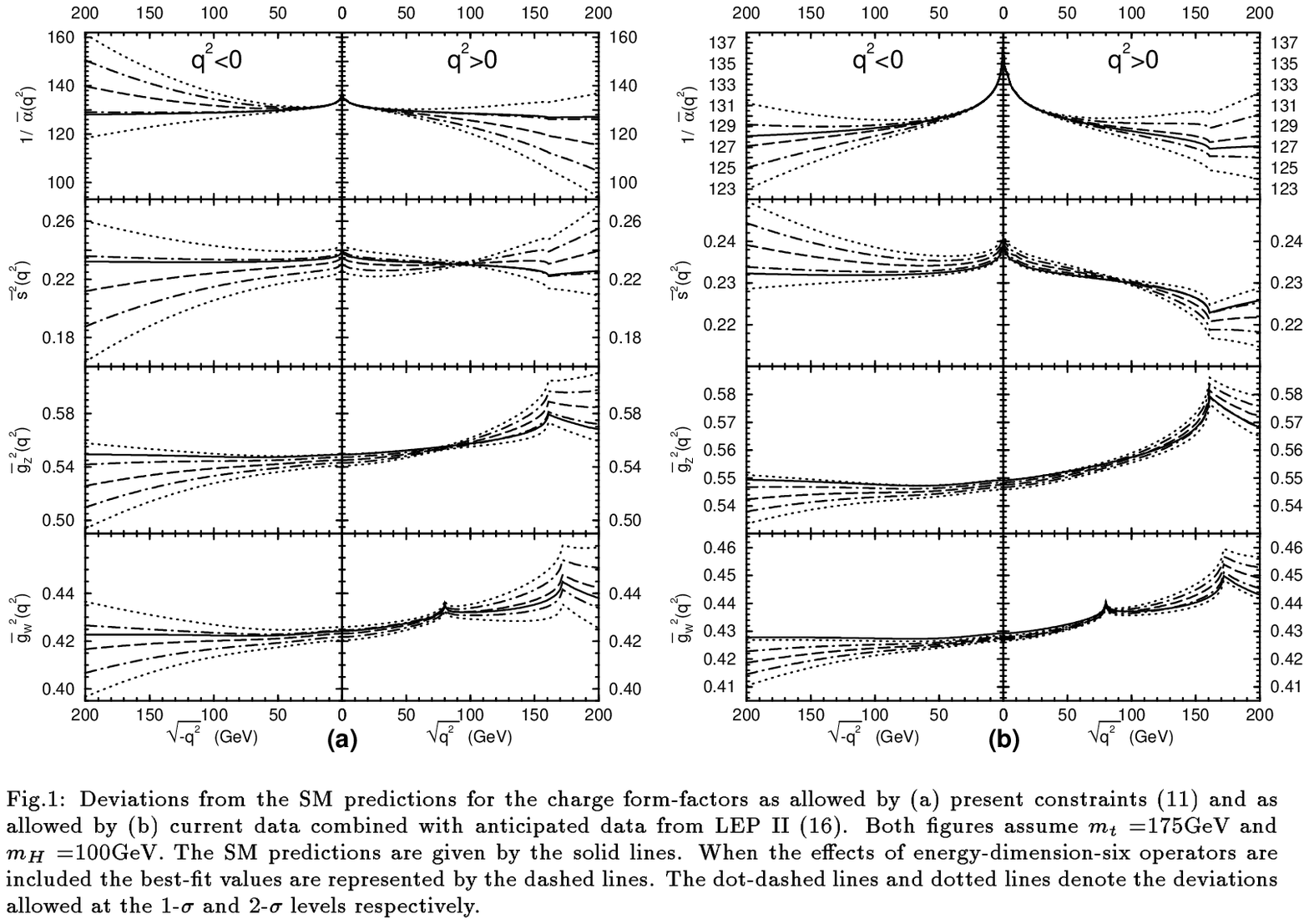,angle=270,height=23cm,silent=0}
\end{figure}
illustrates the possible contribution of the new physics to the running
of the charge form-factors as described by (\ref{corrections})
subject to the constraints of the fit (\ref{fit_now}).
Due to the running with $q^2$, effects which are small at low-energies may
be appreciably enhanced at higher energies.  Alternatively, experiments at
higher $q^2$ are more sensitive to ${\cal O}_{DW}$ and ${\cal O}_{DB}$.
Examine, for instance, the running of
$1/\overline{\alpha}(q^2)$.
Near $q^2=0$, where a direct precision measurement exists,
little deviation is possible.  However, for $q^2=m_Z^2$, the possible
deviations
are fairly large, and for energies relevant to LEP II the constraints are quite
poor.  Similar comments apply to the remaining form-factors.
In particular $\overline{s}^2(q^2)$ is very well constrained at $q^2=m_Z^2$ and
fairly well measured at $q^2=0$.  Away from these points the constraints are
poor.  This behavior is also exhibited by $\overline{g}_Z^2(q^2)$, and
$\overline{g}_W^2(q^2)$ is directly constrained only at $q^2 = 0$.


It is possible to improve these constraints via the increased
precision of existing measurements.  An alternative method is via measurements
at different energy scales.  In the latter
case the absence of high precision may be
compensated by enhancements proportional to $q^2/\Lambda^2$. Those effects of
the new physics which are proportional to $q^2/\Lambda^2$ decouple at low
energies, but, in some cases, at high energies ($|q^2| \geq v^2$) they may
be more important than non-decoupling effects.

The operators ${\cal O}_{DW}$ and ${\cal O}_{BW}$ also contribute to WWZ and
WW$\gamma$ vertices, hence one might hope to achieve additional constraints by
studying processes such as W-boson pair production at LEP II.  However, three
additional operators which are very poorly constrained at present also
contribute to these same vertices\cite{HISZnew}.  Therefore, without making
additional assumptions, this line of analysis is not useful for the current
discussion.

{\bf TRISTAN:}
Consider the measurement of $\overline{\alpha}(q^2)$ at
TRISTAN\cite{tristan} which is especially interesting because it is the first
precision measurement of this quantity near $q^2 = m_Z^2$.
\begin{equation}\label{alpha_tristan}
\overline{\alpha}^{\, -1}\Big((58\rm{GeV})^2\Big)
			= 128.9 \pm 0.6(\rm{stat.}) \pm 2.0(\rm{syst.})\;.
\end{equation}
The combined error of 1.6\% is systematics dominated. The largest
contribution, arising from the measurement of the luminosity, is expected to
decrease further.  Currently the results of the fit (\ref{fit_now}) and the
TRISTAN measurement (\ref{alpha_tristan}) are consistent, but with a
reduced error
this measurement will provide an additional constraint on the coefficients of
the dimension-six operators.  We assume that the combined error of
(\ref{alpha_tristan}) can be reduced to $\approx\pm 0.8$.  For the central
value
of the improved measurement we adopt the SM prediction,
$\overline{\alpha}^{-1}((58{\rm GeV})^2) = 129.46$, and repeat the
analysis with the following results:
\begin{eqnarray}
\label{fit_tristan}
\left.
\begin{array}{lll}
f_{DW} & = & -0.46 + 0.013 \ln x_H - 0.14 x_t \pm 0.61 \\[1mm]
f_{DB} & = & -4.2 \pm 7.4 \\[1mm]
f_{BW} & = & 1.4  + 0.090 \ln x_H \pm 1.8 \\[1mm]
f_{\Phi,1} & = & 0.14 - 0.030 \ln x_H + 0.35 x_t  \pm 0.13
\end{array}
\right.
\makebox[2mm]{}
\left(
\begin{array}{lddd}
\dec 1.   & \dec $-$0.270 & \dec    0.015 & \dec $-$0.447  \\[1mm]
          & \dec    1.    & \dec $-$0.955 & \dec $-$0.665  \\[1mm]
          &               & \dec    1.    & \dec    0.847  \\[1mm]
          &               &               & \dec    1.
\end{array} \right)
\end{eqnarray}
The measurement of $\overline{\alpha}(q^2)$ is a direct constraint upon
$f_{DW}+f_{DB}$, but the primary effect is a reduced one-sigma limit for
$f_{DB}$.  Because of the strong correlations this in turn leads to
smaller errors for $f_{BW}$ and $f_{\Phi,1}$.

The fit (\ref{fit_tristan}) does not qualitatively alter the running of the
charge form-factors from the scenario illustrated in Fig.~1(a), but
numerically the results are
non-negligible.  The allowed deviations in $\overline{\alpha}^{\, -1}(q^2)$
have,
at $\sqrt{q^2}=$200GeV, been reduced by a factor of two.
The corresponding deviations in
$\overline{s}^{2}(q^2)$ and $\overline{g}_Z^2(q^2)$ are reduced by a similar
factor, but there is minimal improvement for $\overline{g}_W^2(q^2)$.

{\bf LEP II:}
Further improvement will likely await results from LEP II.  We assume that
500pb$^{-1}$ of data will be collected at $\sqrt{s}=$175GeV and,
assuming errors dominated by statistics, perform a
fit using the following observables: $\sigma (e^+e^- \rightarrow \mu^+\mu^- )$,
$\sigma (e^+e^- \rightarrow {\rm hadrons})$ and the forward-backward
asymmetries $A^\mu_{FB}$, $A^b_{FB}$ and $A^c_{FB}$.
Additionally we should expect an improved measurement of $m_W$.  We employ
$\Delta m_W \approx 50$MeV.  In the
absence of actual experimental results we choose for the central values of the
hypothetical data the predictions of the SM.  This will bias the central
values of the $f_i$'s which we obtain in our fit, but the estimation of the
errors and correlations will provide useful information.

Combining the hypothetical LEP II data described above with the current data
we repeat the $\chi^2$
analysis and summarize the results:
\begin{eqnarray}
\label{fit_lep2}
\left.
\begin{array}{lll}
f_{DW} & = & -0.52 + 0.043 \ln x_H - 0.43 x_t \pm 0.27 \\[1mm]
f_{DB} & = & 1.5 + 0.10 \ln x_H - 0.52 x_t  \pm 2.1 \\[1mm]
f_{BW} & = & 0.07  + 0.055 \ln x_H + 0.41 x_t \pm 0.54 \\[1mm]
f_{\Phi,1} & = & 0.09 - 0.033 \ln x_H + 0.39 x_t  \pm 0.05
\end{array}
\right.
\makebox[1mm]{}
\left(
\begin{array}{lddd}
\dec 1.   & \dec $-$0.633 & \dec    0.317 & \dec $-$0.089  \\[1mm]
          & \dec    1.    & \dec $-$0.819 & \dec $-$0.360  \\[1mm]
          &               & \dec    1.    & \dec    0.778  \\[1mm]
          &               &               & \dec    1.
\end{array} \right)
\end{eqnarray}
The errors are greatly reduced for all four coefficients.  The correlation
matrix now exhibits a strong $f_{DW}$-$f_{DB}$ anti-correlation, but the
$f_{DB}$-$f_{BW}$-$f_{\Phi,1}$ correlations are weakened.   These new
features are the result of improved knowledge of $\overline{\alpha}^{-1}(q^2)$
well above the Z peak, especially through the measurement of
$\sigma (e^+e^- \rightarrow \mu^+\mu^- )$.
We present Fig.~1(b).  Note that Fig.~1(a) and Fig.~1(b) employ different
scales for the vertical axes.
At 175GeV it is possible to fit $\overline{\alpha}^{-1}$,
$\overline{s}^{2}$, $\overline{g}_Z^{2}$ and  $\overline{g}_W^{2}$ with a
precision of 1.2\%, 0.9\%, 0.5\% and 0.7\% respectively.

{\bf NLC:}
Finally we consider a Next Linear Collider (NLC) with an integrated luminosity
of 50fb$^{-1}$ at $\sqrt{s}=$500GeV.  We assume that the errors in
$R_h$, $A^\mu_{FB}$, $A^\mu_{b}$ and $A^\mu_{c}$ are dominated by
statistics, and hence can be measured to
within approximately one percent.  For the measurement of
$\sigma (e^+e^- \rightarrow \mu^+\mu^- )$ we assume that systematics are
relevant and estimate a three percent error.  We choose for the central values
the predictions of the SM.  The combination of the current data and the NLC
data yields
\begin{eqnarray}
\label{fit_nlc}
\makebox[-1.5mm]{}
\left.
\begin{array}{lll}
f_{DW} & = & -0.04 + 0.0092 \ln x_H - 0.082 x_t \pm 0.06 \\[1mm]
f_{DB} & = & -0.04 - 0.0087 \ln x_H \pm 0.22 \\[1mm]
f_{BW} & = & 0.16  + 0.099 \ln x_H \pm 0.27 \\[1mm]
f_{\Phi,1} & = & -0.08 - 0.030 \ln x_H + 0.36 x_t  \pm 0.04
\end{array}
\right.
\left(
\begin{array}{lddd}
\dec 1.   & \dec $-$0.547 & \dec $-$0.010 & \dec $-$0.108  \\[1mm]
          & \dec    1.    & \dec $-$0.172 & \dec $-$0.013  \\[1mm]
          &               & \dec    1.    & \dec    0.864  \\[1mm]
          &               &               & \dec    1.
\end{array} \right)
\end{eqnarray}
All four coefficients are now measured to the same level of precision, and the
strong anti-correlations between $f_{DB}$ and the remaining coefficients
disappear.



{\bf Acknowledgements:}
A portion of this research was done while R.~Szalapski was in attendance of the
Japan Summer Institute which is supported by the National Science Foundation of
the United States.
Productive discussions with D.~Zeppenfeld
and M.C.~Gonzalez-Garcia are gratefully
acknowledged.


\renewenvironment{thebibliography}[1]
{\small
 \begin{list}{[\arabic{enumi}]}
 {\usecounter{enumi} \setlength{\parsep}{1pt}
  \setlength{\itemsep}{1pt} \settowidth{\labelwidth}{[#1]}
  \settowidth{\leftmargin}{[100]}
  \sloppy}}
 {\end{list}}


\end{document}